\documentclass[a4paper,11pt]{article}
\pdfoutput=1 

\usepackage{jinstpub} 
\usepackage{graphicx}
\usepackage{xcolor}
\usepackage{multirow}

\usepackage{lineno}

\title{\boldmath Detector optimization to reduce the cosmogenic neutron backgrounds in the TAO experiment}

\author[a,b]{Ruhui Li}
\author[a]{, Guofu Cao}
\author[a,b]{, Jun Cao}
\author[a]{, Yichen Li}
\author[a,b]{, Yifang Wang}
\author[a]{, Zhimin Wang}
\author[a]{and Liang Zhan}

\affiliation[a]{Institute of High Energy Physics, Beijing 100049, China}
\affiliation[b]{University of Chinese Academy of Sciences, Beijing 100049, China}

\emailAdd{lirh@ihep.ac.cn}
\emailAdd{wangzhm@ihep.ac.cn}
\abstract{
Short-baseline reactor antineutrino experiments with shallow overburden usually have large cosmogenic neutron backgrounds. The Taishan Antineutrino Observatory (TAO) is a ton-level liquid scintillator detector located at about 30\,m from a core of the Taishan Nuclear Power Plant. It will measure the reactor antineutrino spectrum with high precision and high energy resolution to provide a reference spectrum for JUNO and other reactor antineutrino experiments, and provide a benchmark measurement to test nuclear databases. Background is one of the critical concerns of TAO since the overburden is just 10~meter-water-equivalent. The cosmogenic neutron background was estimated to be $\sim$10\% of signals~\cite{TAO-CDR}. With detailed Monte Carlo simulations, we propose several measures in this work to reduce the neutron backgrounds, including doping Gadolinium in the buffer liquid, adding a polyethylene layer above the bottom lead shield, and optimization of the veto strategy. With these improvements, the neutron background-to-signal ratio can be reduced to $\sim$2\%, and might be further suppressed with pulse shape discrimination.
}


\arxivnumber{2206.01112} 

\begin{document}
\maketitle
\flushbottom

\section{Introduction}
\label{sec:intro}

Short-baseline reactor antineutrino experiments at several to tens of meters from the reactor core usually have very shallow or zero overburden. With very few exceptions, they detect reactor antineutrinos of energy 1.8--10~MeV with hydrogen-rich liquid scintillator (LS) via Inverse Beta Decay (IBD) to search for non-standard neutrino oscillation, measure reactor neutrino spectra, check the reactor antineutrino anomaly, or perform non-proliferation monitoring of the reactor, etc. Due to the shallow overburden, cosmogenic neutron backgrounds are serious backgrounds in these experiments. For example, the ratio of IBD signals to cosmogenic background events is estimated to be 1.37 in PROSPECT~\cite{PROSPECT:2020sxr}, 1 in  STEREO~\cite{STEREO-Allemandou_2018}, 0.5 in  NEUTRINO-4~\cite{neutrino-4-PhysRevD.104.032003}, 1 in Solid~\cite{solid-Abreu_2017}, and 0.05-0.2 in  MINER~\cite{MINER-AGNOLET201753}, respectively. NEOS has about 17\% cosmogenic backgrounds in the IBD sample before applying pulse shape discrimination and about 5\% afterwards~\cite{NEOS-PhysRevLett.118.121802}. For DANSS, the cosmogenic backgrounds constitute 2.7\% of the IBD rate with a relatively large overburden of 50 meter-water-equivalent~\cite{DANSS-ALEKSEEV201856}.

The IBD interaction, $\bar{\nu}_e+p\to n+e^+$, is the typical channel to detect the reactor antineutrino $\overline{\nu}_{e}$ in the few-MeV range with LS detectors\cite{firstfind}. In this reaction, the electron antineutrino interacts with a proton ($p$) in the LS, creating a positron ($e^+$) and a neutron ($n$). The $e^+$ quickly deposits its energy and annihilates into gammas, giving a prompt signal. The neutron is thermalized in the detector and captured by a nucleus, giving a delayed signal. The correlation of the prompt and delayed signals is a powerful signature to tag the reactor antineutrino signal and suppress possible backgrounds. Backgrounds are usually classified into correlated and uncorrelated backgrounds. A correlated background refers to a pair of prompt and delayed events coming from the same source, while an uncorrelated background refers to the prompt and delayed events coming from different origins but falling into the time window accidentally, thus is also called an accidental background. An energetic neutron entering the detector can form a fast-neutron background by recoiling off a proton before being captured. Multiple neutrons produced by the same muon may be captured in the detector and form a prompt-delay pair in the IBD event selection window, referred as a double-neutron background. There are other correlated cosmogenic neutron backgrounds, e.g. at least one of the prompt and delayed signals comes from a gamma released by neutron capture on outer detector material. Uncorrelated backgrounds and other correlated backgrounds, such as long-lived cosmogenic isotopes, are also important backgrounds, but less significant for a shallow-overburden experiment and will not be discussed in this work.

The Taishan Antineutrino Observatory (TAO, also known as JUNO-TAO)~\cite{TAO-CDR} is a satellite experiment of the Jiangmen Underground Neutrino Observatory (JUNO)~\cite{JUNO-CDR,JUNO-yellow-book-2016,JUNO-detector}. It consists of a ton-level liquid scintillator (LS) detector at around 30 meters from a reactor core of the Taishan Nuclear Power Plant in China. TAO is designed to have an energy resolution of better than 2\%/$\sqrt{E[{\rm MeV}]}$. It will measure the reactor neutrino spectrum with high precision and high energy resolution to provide a reference spectrum for JUNO and to provide a benchmark measurement to test nuclear databases. The spectral shape uncertainty is expected to be measured to $<1$\% in most of the energy region. Therefore, the uncertainty from background subtraction should be less than 1\%.


In the TAO Conceptual Design Report (CDR)~\cite{TAO-CDR}, the major backgrounds of TAO were estimated to be $\sim$10\% for the accidental coincidences, $\sim$10\% for the cosmogenic neutron backgrounds, and $\sim$2.5\% for the long-lived isotopes $^8$He/$^9$Li. The accidental backgrounds can be accurately subtracted by measuring singles {\it in situ}, thus have little impact on the spectral shape precision. To further reduce the cosmogenic neutron backgrounds, in this report we study the formation of the neutron backgrounds, optimize the detector, and improve the veto strategy with detailed Monte Carlo simulations. We present a brief description of the TAO detector design~\cite{TAO-CDR} in Sec.~\ref{sec:taodesign}, and describe the cosmogenic neutron background simulation in Sec.~\ref{sec:simulation}. Several measures to further reduce the neutron backgrounds are studied in Sec.~\ref{sec:optimization}, followed by a summary.

\section{TAO detector design}
\label{sec:taodesign}

The TAO detector will be installed in a basement at 9.6~m underground, outside of the concrete containment shell of the reactor core. The distance to the center of the core is about 30 m.  Shielded by the roof and the floor plates of the building equivalent to an overburden of $\sim$10 meters-water-equivalent on top, the muon rate and cosmogenic neutron rate in the laboratory are measured to be 1/3 of those on the surface.

\begin{figure}[htbp]
    \centering 
    \includegraphics[width=0.7\textwidth,origin=c,clip]{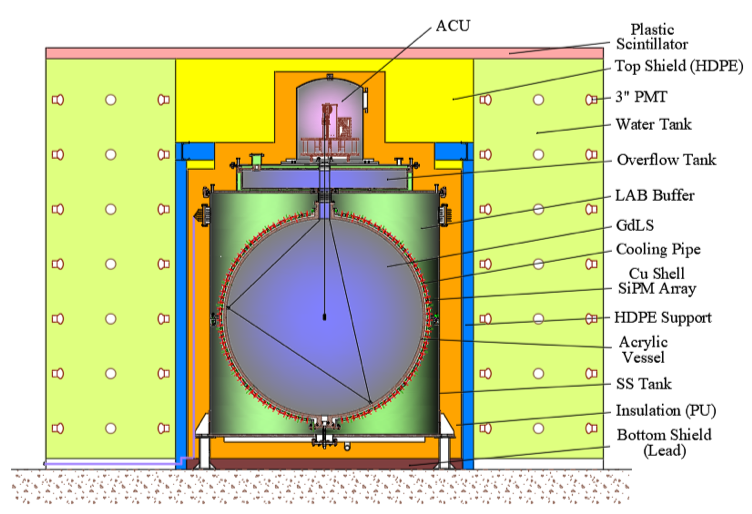}
    \qquad
    \caption{Schematic view of the TAO detector, which consists of a Central Detector (CD) and an outer shielding and veto system. The CD consists of 2.8~ton Gadolinium-doped liquid scintillator (GdLS) filled in a spherical acrylic vessel and viewed by $10~{\rm m}^2$ SiPM, a spherical Copper Shell that supports the SiPMs, 3.45~ton buffer liquid, and a cylindrical stainless steel tank insulated with 15-cm thick melamine foam. The outer shielding includes 1.2~m  water in the surrounding tanks, 1~m High Density Polyethylene (HDPE) on the top, and 10~cm lead at the bottom. The water tanks, instrumented with Photomultipliers, and the Plastic Scintillator on the top comprise the active muon veto system. Taken from TAO CDR~\cite{TAO-CDR}.}
    \label{fig:tao}
\end{figure}

The antineutrino detector, called Central Detector (CD), is shielded with water tanks on the sides and High Density Polyethylene (HDPE) on the top, as shown in Fig.~\ref{fig:tao}. The antineutrino target is 2.8~ton Gadolinium-doped liquid scintillator (GdLS) contained in a spherical acrylic vessel of 1.8~m in diameter. The Gd content in GdLS is 0.1\% by weight. The fiducial mass is 1~ton after the event vertex selection of $>$25~cm from the acrylic wall. The GdLS is viewed by 10~m$^2$ Silicon Photomultipliers (SiPMs) of photon detection efficiency $\sim$50\%, implemented as an array of 4100 SiPM tiles, each of $5\time5$~cm$^2$ in dimention. The SiPMs are installed on a spherical Copper Shell wrapping the acrylic vessel. The clearance between the inner wall of the Copper Shell and the outer wall the acrylic vessel is 21~mm. The Copper Shell is submerged in Linear Alkylbenzene (LAB) as the buffer liquid, contained in a stainless steel outer tank of 2.1~m in diameter  and 2.2~m in height. To reduce the dark noise of the SiPMs, the whole CD will be operated at -50 $^{\circ}$C.

The water tanks provide 1.2-m water shielding from the sides of the CD. Three water tanks surround the CD and form a dodecagon barrel. A 10-cm layer of lead bricks will be installed at the bottom of the CD, and about 1\,m HDPE will be placed on the top. The main role of the lead bricks is to shield gammas from radioactivity, and 10~cm thickness is enough. With this shielding, the rate of the radioactivity singles from the surroundings and detector material is estimated to be $<$100~Hz.

The water tanks are instrumented with a total of 300 3-inch Photomultipliers (PMT) to form an active Cherenkov muon detector. The muon detection efficiency is estimated from simulation to be $>$99\%. Four layers of instrumented plastic scintillator strips will cover the top of the CD, with an efficiency of $>$98\% for downward-going cosmic muons. The muon rate passing the water tanks and top plastic scintillators is about 3300~Hz.

The estimated antineutrino and background event rates in the TAO detector presented in the TAO CDR~\cite{TAO-CDR} are shown in Table~\ref{tab:background}. Comparing to the CDR, the present detector design has been slightly improved, e.g. the geometry of the water tanks and layout of the top plastic scintillators. Simulations are also improved by the TAO team. However, all rates calculated in the new simulations are similar to those in Table~\ref{tab:background}.

\begin{table}
\setlength{\belowcaptionskip}{5pt}
\begin{center}
\caption{Summary of the IBD signal and background simulation results in the TAO detector, taken from the TAO CDR~\cite{TAO-CDR}. \label{tab:background}}
\begin{tabular}{r l}
  \hline\hline
   IBD signal & \mbox{ }\mbox{ } 2000~events/day \\
  Muon rate & \mbox{ }\mbox{ } 70~Hz/m$^2$ \\
  Cosmogenic neutron backgrounds & $< 200$~events/day \\
  Singles from radioactivity & $< 100$~Hz \\
  Accidental background rate & $< 190$~events/day \\
  $^{8}$He/$^{9}$Li background rate & $\sim 54$~events/day \\
  \hline
\end{tabular}
\end{center}
\end{table}

\section{Cosmogenic background simulation}
\label{sec:simulation}

\subsection{Simulation software}
A Geant4\cite{geant4-AGOSTINELLI2003250}-based detector simulation software has been developed for the detector design optimization and background study, and is described in the TAO CDR~\cite{TAO-CDR}. The GdLS response is modelled similar to the Daya Bay experiment. For simplicity, optical simulation, electronics simulation, and reconstruction have not been included in this optimization study. The event energy is from the LS-quenched energy deposition in the detector, and the event vertex is the energy deposition weighted center, which is calculated by the position of every step in Geant4 weighted by its energy deposition of each step. A full simulation would require additional energy and vertex smearing, but we expect that the conclusion of this study using a simplified simulation would not be significantly changed.

The detector geometry is accurately modeled in the simulation, while the laboratory geometry model is simplified and does not represent the actual building structure. The TAO detector is installed in a basement at 9.6\,m underground. In the simulation, the dimension of the laboratory is 10\,m $\times$ 10\,m $\times$ 10\,m, which is surrounded by 20\,m thick rock from the sides and the bottom. The top overburden is the roof of the building and five layers of floor plates. They are modeled in the simulation as six layers of concrete with a thickness of 66\,cm each. The thickness of the concrete has been tuned in the simulation to let the muon rate be consistent with measurements in the laboratory.

\begin{figure}[htb]
    \centering

    \subfigure[]{\includegraphics[width=0.38\linewidth]{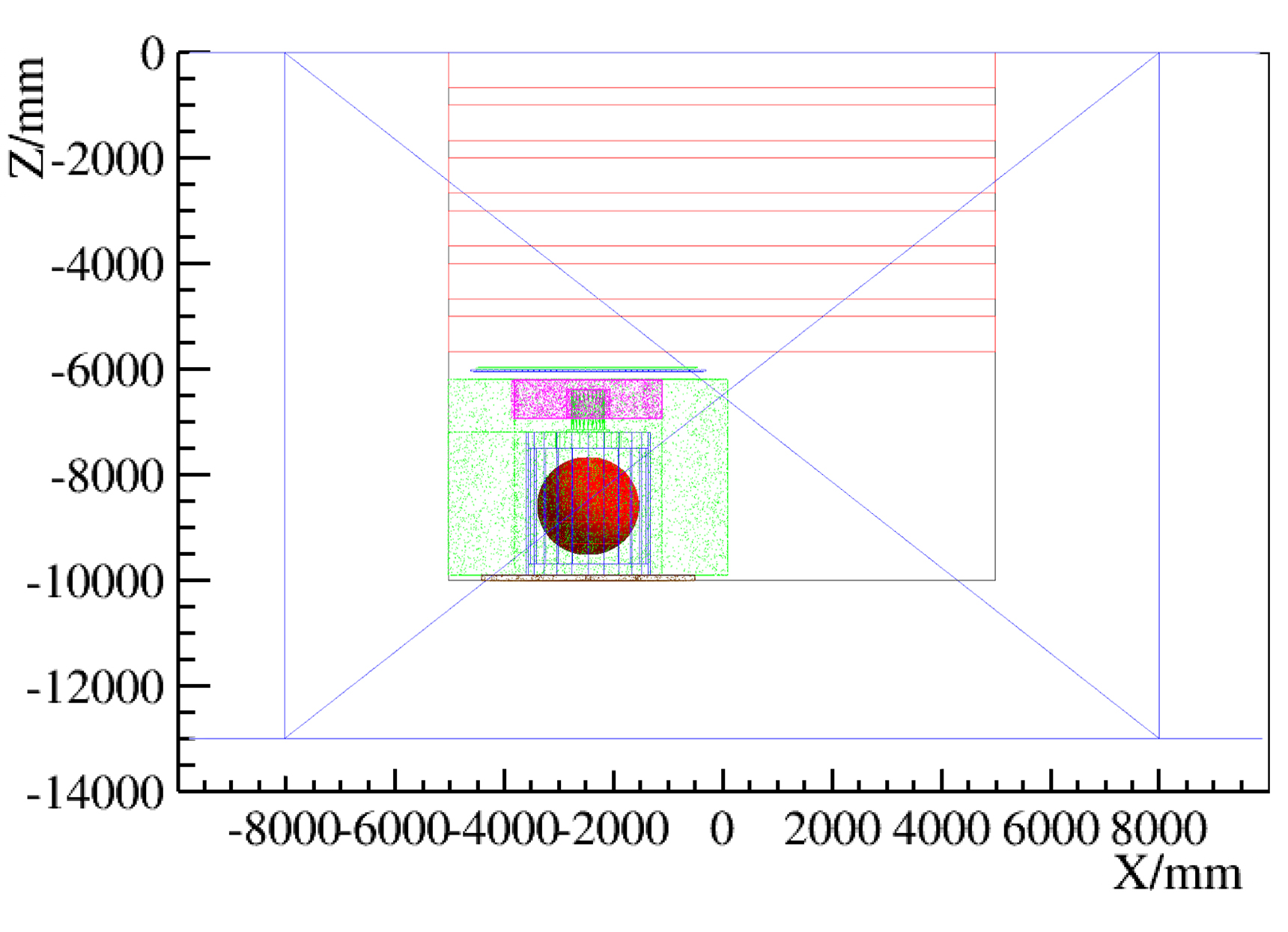}}
	\subfigure[]{\includegraphics[width=0.45\linewidth]{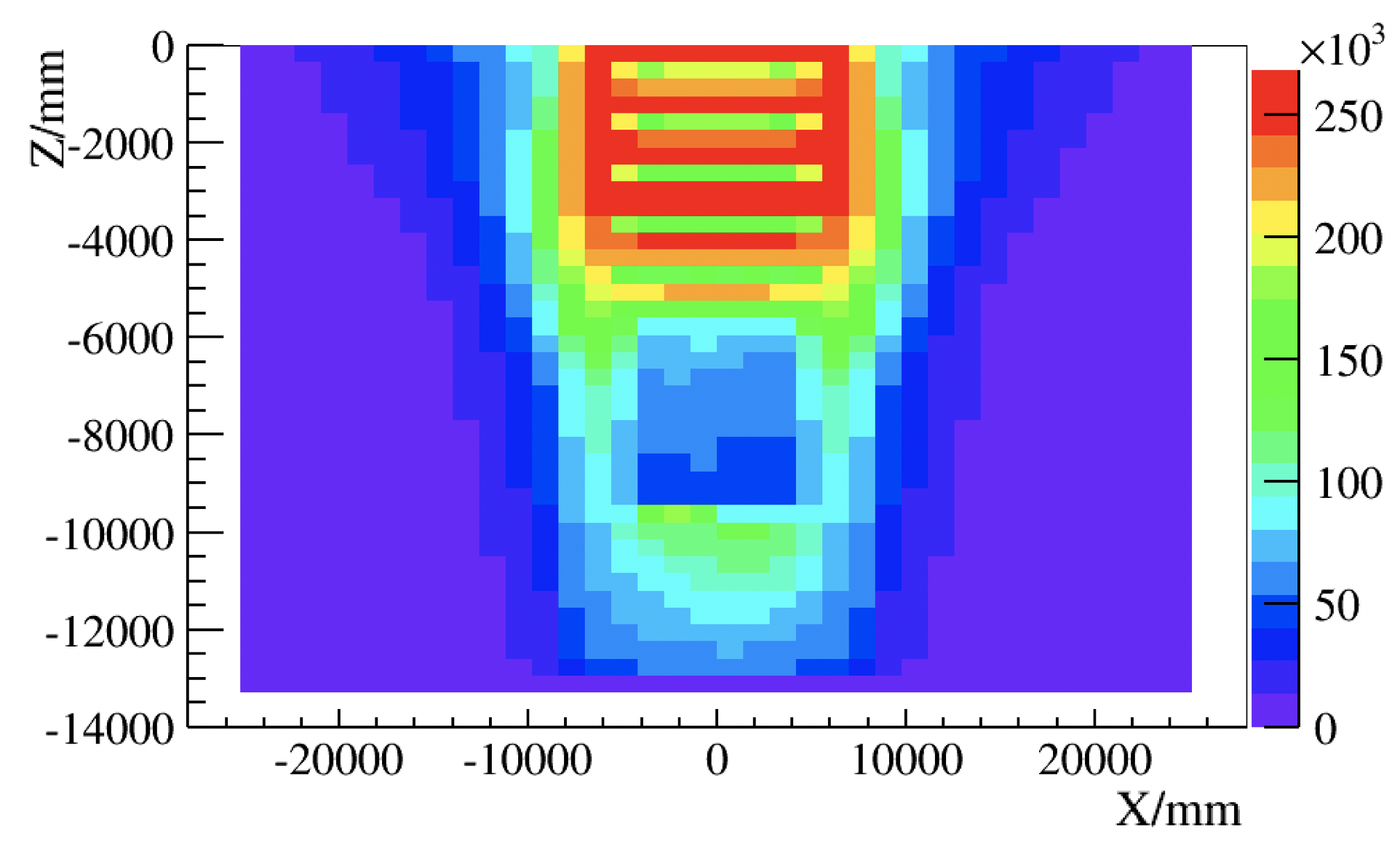}}
    \caption{(a) Geometry in Geant4. The detector is located in the lower left corner of the laboratory, which is surrounded by 20\,m rock. On top of the detector, there are six layers of concrete to model the roof and floor plates of the building. (b) Primary neutron production rates at various locations from the simulation of 1400-second muon.}
    \label{fig:sim:neutron:gen}
\end{figure}

Fig.~\ref{fig:sim:neutron:gen}(a) shows the geometry of the laboratory in the simulation. Muons are generated uniformly on a 50\,m$\times$50\,m plane on the ground. The direction and energy information included in the HEPEVT file serve as the generator. Muons with the shortest distance to the experimental hall $>$3~m will be ignored to speed up the simulation, since their secondaries contribute negligible backgrounds in the TAO detector as we found by simulation. Fig.~\ref{fig:sim:neutron:gen}(b) shows the production rate of all primary neutrons obtained in the simulation at various locations, no matter whether they are detected or not. The neutron rate shown in the figure corresponds to a simulation of 1400 seconds data taking. The neutron production rate in the floor plates above the detector is the highest since the vertical muon rate is significantly higher than in other directions due to the lack of rock shielding. The production rates are listed in Table~\ref{tab:neutron:gen} for each detector component, from which we find that High-Z materials such as lead, copper, and stainless steel generate significantly more neutrons per unit mass than light materials.

\begin{table}[htbp]
\centering
\caption{The primary neutron production rates in each detector component. Water, HDPE, and Pb correspond to 1.2\,m thick water shielding on the sides, 1\,m thick HDPE shielding on the top,  and 10\,cm thick lead at the bottom of the CD, respectively. SST is the stainless steel outer tank of the CD. LAB and Cu are the buffer liquid and the Copper Shell in the CD.}
\label{tab:neutron:gen}
\smallskip
\begin{tabular}{c|c|c|c|c|c|c|c}
\hline
 & Water & HDPE & Pb & SST & LAB & Cu & GdLS \\
\hline
neutron rate (Hz)  & 60.8 & 7.4 & 80 & 5.8 & 2.2 & 3.3  & 1.2 \\
\hline
mass (t)  & 56.2 & 5.6 & 13.9 & 1.4 & 4.0 & 1.1 & 2.6\\
\hline
\end{tabular}
\end{table}

\subsection{Neutron background analysis}

The cosmogenic neutron backgrounds are selected with the following criteria, which is the same as the selection criteria for reactor antineutrinos in the TAO CDR. A prompt (delayed) signal candidate requires an energy in the range 0.9--9~MeV (7--9~MeV). The prompt and delayed signal should be correlated in a time windows of 1--100~$\mu$s. No other signal is present in the CD in 100~$\mu$s before the prompt candidate,  100~$\mu$s after the delayed candidate, or between the prompt and delayed candidates. The 100~$\mu$s time correlation window for the prompt and delayed signals is shorter than the typical 200~$\mu$s in the Daya Bay analysis~\cite{dayabay-PhysRevLett.108.171803} because the muon rate in TAO is much higher. A 200~$\mu$s time window will introduce a large loss of the IBD selection efficiency. With these selections, we obtain the cosmogenic neutron backgrounds in the TAO detector to be 4409 per day before applying a muon veto. To study the formation of these backgrounds, we classify the neutron backgrounds into three categories: double-neutron, fast-neutron, and other. A double-neutron background is two neutron captures on nuclei in the detector. A fast-neutron has a prompt signal from the proton recoil and a delayed signal from the neutron capture on Gd after thermalization. All the other backgrounds are combined into the "other" category, such as the prompt or delayed signals from gammas released by neutron capture in rock, Michel electrons, or any other cosmogenic signal correlated accidentally. The double-neutron, fast-neutron, and other backgrounds are 373, 2858, and 1178 per day, respectively, as shown in Table~\ref{tab:nrates}.

\begin{table}[htbp]
\centering
\caption{Cosmogenic neutron background rate before and after a muon veto from simulation. The background rates after veto are further classified into contributions from tagged muons (i.e. detected by the active veto detectors) and untagged muons. }
\label{tab:nrates}
\smallskip
\begin{tabular}{c|c|c|c|c}
\hline
  (/day) & before veto & after veto & tagged muon & untagged muon \\
\hline
Double-neutron & 373 & 309 & 300 & 9 \\
\hline
Fast-neutron     & 2858 & 53 & 0 & 53 \\
\hline
Other                & 1178 & 9 & 0 & 9 \\
\hline
All                    & 4409 & 371 & 300 & 71  \\
\hline
\end{tabular}
\end{table}

To reject the neutron backgrounds, a veto strategy is applied as following.
\begin{itemize}
\item[a)] Outer muon veto: veto the CD by 20~$\mu$s if a muon is detected (aka ``tagged") by the veto detectors, i.e. the water tanks or the top plastic scintillators, but not by the CD;
\item[b)]  CD muon veto: veto the CD by 100~$\mu$s if the muon is also tagged by CD, identified by a CD event with energy $>$0.7~MeV correlated with the veto detectors in 100~ns;
\item[c)] Spallation muon veto: veto the CD by 200~$\mu$s if there are neutrons detected after a muon, identified by a 6-10~MeV signal in 100~$\mu$s (20~$\mu$s) after a CD muon (Outer muon), or there is a $>$10 MeV signal in the CD in 20~$\mu$s after any tagged muon.
\end{itemize}
The neutron capture time in 0.1\% GdLS is about 28~$\mu$s. In experiments with enough overburden, usually at least 200~$\mu$s veto time is chosen to reject neutron backgrounds, e.g. in Daya Bay~\cite{dayabay-PhysRevLett.108.171803}. Due to the high muon rate in the TAO detectors (3300~Hz), it is impossible to apply a long enough veto time window. If a similar 200~$\mu$s veto time is applied, the detector dead time would be more than 60\%. Balancing the detector live time and the background rejection efficiency, a short veto time of 20~$\mu$s is applied for a muon not passing the GdLS, and thus having small probability to produce a neutron background. A longer veto time of 100~$\mu$s is applied for a muon or its daughters passing the GdLS, and thus having higher probability to produce a neutron background. If a neutron or a signal of $>$10 MeV is detected after a muon, most likely a spallation process has happened in or near the GdLS. A long veto time of  200~$\mu$s is applied since the probability of producing a neutron background is high.

The cosmogenic neutron background rates after applying the muon veto cuts are shown in Table~\ref{tab:nrates}. After the muon veto, the remaining backgrounds are reduced to 371 per day, among which there are 309, 53, and 9 for double-neutron, fast-neutron, and other backgrounds. We also separate the contributions from tagged and untagged muons to understand the origin of these backgrounds and optimize the detector design. After selection, 80\% of the neutron backgrounds are double-neutron backgrounds with their parent muons tagged. However, due to the short veto time, multiple neutrons produced by the muon have a large chance to survive the veto and form a double-neutron background. They are discussed in the following section.

\section{Design optimization}
\label{sec:optimization}
\subsection{Adding Gd in LAB buffer}

From Table~\ref{tab:nrates} we know that double-neutron backgrounds dominate. We have further studied the production location of the primary neutrons of these backgrounds, and found that most of them are produced in the Copper Shell, as shown in Table~\ref{tab:genpos}. Out of the total 309 double-neutron backgrounds, 216 are from the Copper Shell, and 34, 13, 4, and 42 are from the lead at the bottom of CD, GdLS in the CD, outside of the detector system, and other detector components, respectively. Outside of the detector system means the rock and the floor plates. Other detector components include the water tanks, HDPE, stainless steel tank, buffer liquid, SiPM, and calibration system, etc.

\begin{table}[htbp]
\begin{center}
\caption{Production location of the primary neutrons for the neutron backgrounds. Cu, Pb, and GdLS labels the Copper Shell, lead at the bottom of the CD, and GdLS in the CD. ``Outside" labels the rock and floor plates outside of the detector. And ``Others" includes all other detector components.
\label{tab:genpos}}
\smallskip
\begin{tabular}{c|c|c|c|c|c|c}
\hline
(per day) & Cu & Pb & GdLS & Outside & Others & All\\
\hline
Double-neutron & 216 & 34 &13 & 4& 42&309\\
\hline
Fast-neutron & 11 & 0&0 & 36& 6&53\\
\hline
\end{tabular}
\end{center}
\end{table}

Neutrons produced in the Copper Shell have a big chance of entering the GdLS because the clearance is only 2~cm. The neutron capture time is 28~$\mu$s in the 0.1\% GdLS and $\sim$200~$\mu$s in the buffer LAB.
However, many neutrons live much longer than 28~$\mu$s before being captured on Gd since neutrons around the boundary of the GdLS could penetrate the boundary back and forth multiple times before capture. These neutrons can easily survive the muon veto cuts. Such neutron backgrounds could be reduced if we also dope Gd into the buffer liquid LAB to shorten the neutron capture time. Furthermore, these long-lived neutrons must be captured near the wall of the acrylic vessel. When a fiducial volume cut of 25~cm is applied, a large fraction of these backgrounds will be rejected. The simulation results are shown in Table~\ref{tab:gd}. After loading 0.1\% Gd in LAB, more than 40\% of the double-neutron backgrounds can be rejected. The fast-neutron backgrounds decrease little because most of them can be rejected easily with a short capture time. After the fiducial volume cut, most of the neutron backgrounds can be rejected, resulting in a neutron background rate of 48 per day, or 2.5\% in the background-to-signal ratio.

After doping 0.1\% Gd in LAB, most neutrons captured in the buffer will release gammas of a total energy of  $\sim$8~MeV, comparing to a 2.2~MeV gamma from the capture on hydrogen without doping. If these gammas enter the GdLS, they might provide a delayed signal to be correlated with other singles and form a coincidence background. We find with simulations that this increase in background is negligible.

\begin{table}[htbp]
\centering
\caption{Comparison of the neutron backgrounds without and with 0.1\% Gd in the buffer LAB, and without and with fiducial volume cut (FV).}
\label{tab:gd}
\smallskip
\begin{tabular}{c|c|c|c|c}
\hline
\multirow{2}{*}{(per day)}&\multicolumn{2}{|c|}{without Gd in LAB}&\multicolumn{2}{|c}{adding 0.1\% Gd in LAB}\\
\cline{2-5}
 & w/o FV &  w/ FV & w/o FV& w/ FV \\
\hline
Double-neutron background & 309&72 &179 &37 \\
\hline
Fast-neutron background &53 &12 &45 &11 \\
\hline
Other &9 &2 & 15&0 \\
\hline
All&371&86&239&48\\
\hline
\end{tabular}
\end{table}

\subsection{Adding HDPE on lead}

In Table~\ref{tab:genpos}, it shows that the lead contributes the second largest part of the neutron background in a single detector component. High-Z material produces much more neutrons than low-Z materials. For example, neutron yields per unit mass in lead, copper, and stainless steel is 12, 6, and 9 times higher than that in liquid scintillator in the TAO detector, although the accurate ratio is limited by statistics and depends on the input muon energy spectrum. However, due to the limited space in the laboratory, it is unlikely to be able to change the shielding material at the CD bottom. Adding more hydrogen-rich material between the bottom lead and the CD could moderate the neutrons and prevent them from entering the GdLS. Simulation with a 10\,cm-thick HDPE layer added shows that the neutron background from lead is reduced by about 75\%.

\subsection{Adding extra HDPE on the top}

Although we have a 3.85\,m height limit for the integrated detector system in the laboratory, there is still fragmented space to put extra HDPE above the top plastic scintillators for additional neutron shielding. Intuitively, adding additional HDPE to moderate neutrons seems an effective way to reduce neutron backgrounds. However, simulations with an additional 60~cm HDPE show that neutron backgrounds originating from the top of the detector are reduced by only a few percent. A possible explanation is that adding HDPE above the top plastic scintillator won’t change the neutron backgrounds originating from the HDPE shielding between the plastic scintillator and the central detector, and the fast-neutron backgrounds originating from the floor plates are not significant in the TAO detector. In this case, additional HDPE above the plastic scintillator won't change the neutron backgrounds too much.

\subsection{Veto strategy optimization}

The default veto time used in above analyses is roughly estimated with the detector live time and the neutron capture time in the detector. To optimize the veto time for a precision measurement of the antineutrino spectrum, we define a figure of merit (FOM) to scan for the best time windows as
\begin{equation}
\textit{FOM} = \frac{1}{N_S}+(\sigma_B \times\frac{N_B}{N_S})^2  \,,
\label{eq:significance}
\end{equation}
where $N_S$ is the number of selected antineutrino signals per day, $N_B$ is the corresponding neutron background events per day, and $\sigma_B$ is the uncertainty from the background subtraction. In the figure of merit, $1/N_S$ represents the statistical uncertainty which dominates the precision of the spectral measurement. There are multiple ways to estimate $\sigma_B$. As the neutron backgrounds can be directly measured during reactor shutdown, here we take this simplest and most conservative approach.
Assuming that the reactor-off time is 90 days in 3 years, we will have a neutron background sample of $90\times N_B$. About 300 energy bins in the whole IBD energy range will be used for the mass hierarchy analysis in JUNO. Taking the same binning in TAO and assuming the energy spectrum of the neutron backgrounds is relatively flat, each energy bin will have $90\times N_B/300$ neutron background events. Then, the shape uncertainty due to background subtraction is $\sigma_B = 1/\sqrt{90\times N_B/300}$ by statistics of the reactor-off background measurement in 3 years. With a longer veto time, neutron backgrounds $N_B$ reduces, but the detector live time also reduces thus the statistical uncertainty increases. A smaller figure of merit corresponds to better spectrum measurement precision.

\begin{figure}[htbp]
    \centering 
    \includegraphics[width=0.7\textwidth,origin=c,clip]{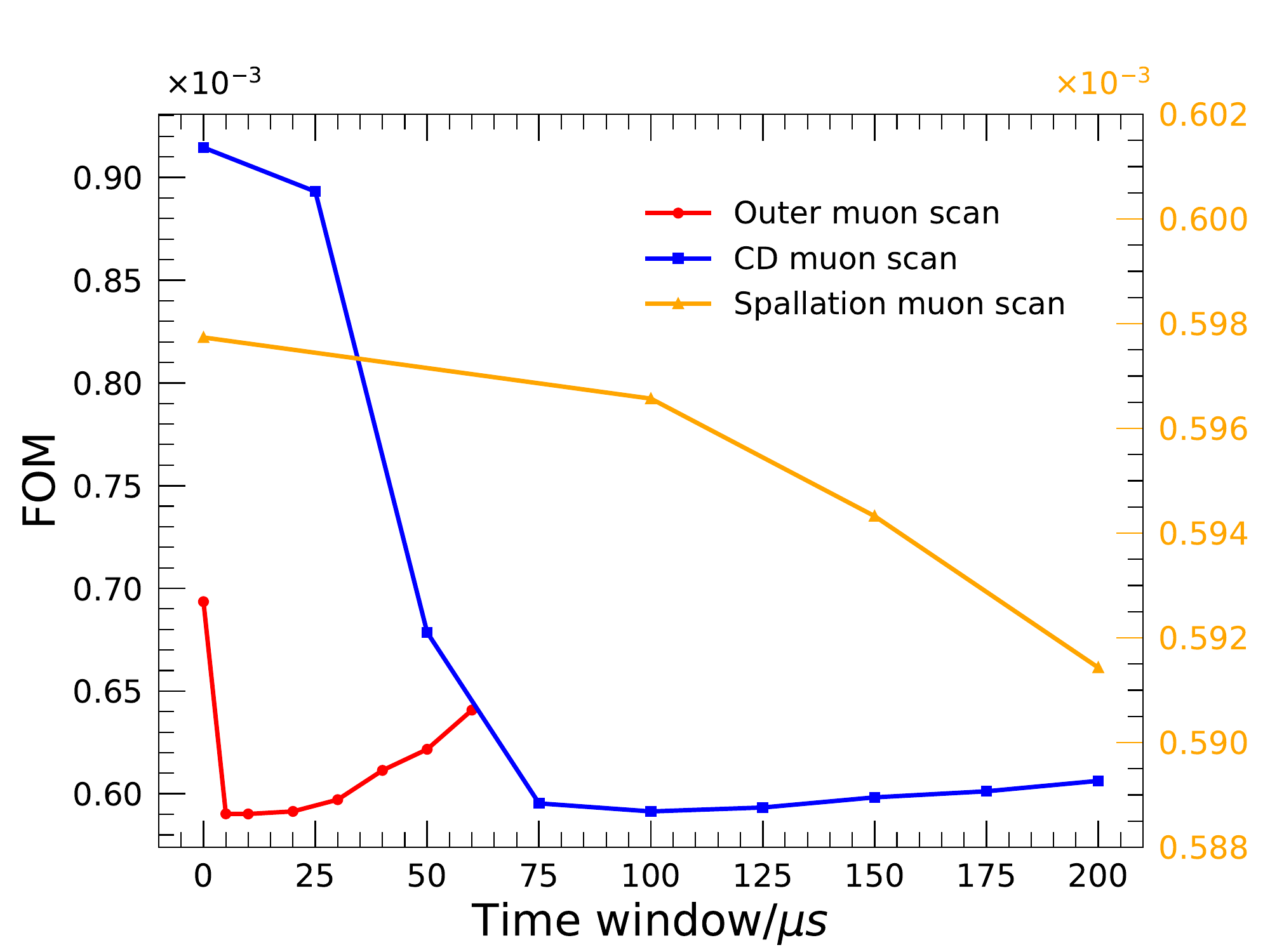}
    \qquad
    \caption{Figure of merit scan for the veto time for the Outer muon (red line), CD muon (blue line), and Spallation muon (orange line).}
    \label{fig:sim:Veto:time:opt}
\end{figure}

The scan results are shown in Fig.~\ref{fig:sim:Veto:time:opt}. The red line shows the optimization of the Outer muon veto time. The difference among the figures of merit corresponding to the 5, 10, and 20~$\mu$s veto time is negligible. The increase for 0 or larger than 20~$\mu$s veto time is moderate. It reflects the fact that the neutron backgrounds from the Outer muons are not significant. A veto time of 20~$\mu$s  for the Outer muon veto could be both safe and optimal.
The blue line shows the optimization of the CD muon veto time. For shorter than 70~$\mu$s, the figure of merit increases fast, indicating that a muon passing the CD or having daughters passing CD has large chance to produce a neutron background. A veto time of 100~$\mu$s is still preferred. The orange line shows the optimization of the Spallation muon veto time. The figure of merit is shown on the right vertical axis. The variation is relatively small comparing to the Outer muon and CD muon. Since the rate of the spallation muon is much lower than other background sources we keep the 200~$\mu$s veto time for the Spallation muon. 

In the current optimization, we have assumed a conservative estimation of the background uncertainty $\sigma_b$ with a direct measurement with reactor-off data. With real data, the strategy could be updated with the actual observed neutron background $N_b$ and the estimated background uncertainty $\sigma_b$.

\subsection{Cosmogenic neutron backgrounds after optimization}

With the optimization studied above, including adding 0.1\% gadolinium in the buffer LAB, adding 10-cm HDPE between the lead shielding and the CD, and using the optimal veto time, the cosmogenic neutron background rates are shown in Table~\ref{tab:bkg:final}. Adding gadolinium in the buffer LAB could significantly reduce the double-neutron backgrounds, especially that from the Copper Shell adjacent to the GdLS. Adding HDPE above the lead shielding could suppress the neutron backgrounds from the lead. Optimization of the veto strategy could better balance between the detector live time and the background veto efficiency. The fiducial volume cut is very effective since most remaining double-neutron backgrounds are on the edge of the GdLS. We expect 44 cosmogenic neutron backgrounds per day after the fiducial volume cut. Among the 11 neutron backgrounds from untagged muons, 10 are fast-neutrons, which could be further rejected using the pulse shape discrimination of the liquid scintillator detector with 80\% efficiency from a preliminary study. Therefore, with the detector optimization and the optimal veto strategy, we could achieve a background-to-signal ratio of 2\% for the cosmogenic neutron backgrounds, while the detector dead time due to the muon veto is about 8.1\%  and the antineutrino rate is 1838 per day. The dead time consists of 4.8\%, 2.8\%, and 0.5\%  from the Outer muon veto, CD muon veto, and Spallation muon veto, respectively.

\begin{table}[htbp]
\centering
\caption{Cosmogenic neutron background rate with the optimized detector design and veto strategy. Both rates without and with fiducial volume cut (FV) are shown.}
\label{tab:bkg:final}
\smallskip
\begin{tabular}{c|c|c|c|c|c}
\hline
 (per day) & \multicolumn{3}{|c|}{From tagged muon} & \multirow{2}{*}{From untagged muon} & \multirow{2}{*}{Sum} \\ \cline{2-4}
  & double-neutron &fast neutron& other &  &  \\
\hline
without FV & 137 & 5&2 & 60 &  204\\
\hline
With FV & 31 & 2 &0  & 11 & 44 \\
\hline
\end{tabular}
\end{table}

\section{Summary}
\label{sec:summary}
Cosmogenic neutron backgrounds are the most serious backgrounds in a reactor antineutrino experiment at shallow overburden. The Taishan Antineutrino Observatory is designed to have a background-to-signal ratio of $<$10\% for these backgrounds. We have studied the neutron backgrounds in detail with Monte Carlo simulations. We have found that the dominant background is the double-neutron background with primary neutrons originated from the Copper Shell, which is very close to the neutrino target GdLS. By adding 0.1\% gadolinium in the buffer liquid LAB, the neutron capture time is shortened, and more than 40\% of the double-neutrons are rejected. Adding 10-cm HDPE between the bottom lead shielding and the CD could moderate the neutrons originating from lead and reject 75\% of them. A fiducial volume cut of 25 cm is very effective to reject the double-neutron backgrounds since most of them are on the edge of the GdLS. Adding additional HDPE shielding on top of the top plastic scintillator is not very effective. Combined with optimal veto time, we could achieve a background-to-signal ratio of 2\% for the cosmogenic neutron backgrounds, which is significantly better than the design value in the TAO Conceptual Design Report.

\acknowledgments
This work was supported partially by the National Natural Science Foundation of China (Grant No. 11875282 and 12022505), the Strategic Priority Research Program of the Chinese Academy of Sciences (Grant No. XDA10011200), the joint RSF-NSFC project under Grant No. 12061131008, the CAS Center for Excellence in Particle Physics, and the Youth Innovation Promotion Association of CAS.

\bibliographystyle{unsrtnat}
\bibliography{TAO_nbkg_sim}
\end{document}